\documentclass{article}
\usepackage{amsmath}
\usepackage{amssymb,amsthm}
\usepackage{pdfsync}
\usepackage[english]{babel}

\renewcommand{\d}{\mathrm{d}}
\newcommand{\bt}{\boldsymbol{t}}

\newcommand{\bb}{\boldsymbol{\beta}}

\newtheorem{pro}{Proposition}

\begin{document}

\title{ Singular sectors of the 1-layer Benney and dToda systems and their
interrelations. }

\author{B. Konopelchenko $^{1}$, L. Mart\'{\i}nez Alonso$^{2}$ and E. Medina$^{3}$
\\
\emph{ $^1$ Dipartimento di Fisica, Universit\'a del Salento and Sezione INFN}
\\ {\emph 73100 Lecce, Italy}\\
\emph{$^2$ Departamento de F\'{\i}sica Te\'orica II, Universidad
Complutense}\\
\emph{E28040 Madrid, Spain}\\
\emph{$^3$ Departamento de Matem\'aticas, Universidad de C\'adiz}\\
\emph{E11510 Puerto Real, C\'adiz, Spain}
}

\maketitle

\begin{abstract}

 Complete description of the singular sectors of the 1-layer Benney system ( classical long wave equation)
and dToda system is presented. Associated Euler-Poisson-Darboux
equations E(1/2,1/2) and E(-1/2,-1/2) are the main tool in the
analysis. A complete list of solutions of the 1-layer Benney
system depending on two parameters and belonging to the singular
sector is given. Relation between Euler-Poisson-Darboux equations
$E(\varepsilon,\varepsilon)$ with opposite sign of $\varepsilon$
is discussed.
\end{abstract}

\maketitle
\section{Introduction}

The 1-layer Benney system (classical long wave equation)
\begin{equation}\label{ben}
\begin{cases}
u_t\,+\,u\,u_x\,+\,v_x\,=\,0,\\ \\
v_t\,+\,(u\,v)_x\,=\,0
\end{cases}
\end{equation}
and dToda equation $v_{xx}=(log v)_{tt}$ or equivalently the
system
\begin{equation}\label{toda2}
\begin{cases}
u_t\,+\,v_x\,=\,0,\\ \\
v_t\,+v u_x\,=\,0
\end{cases}
\end{equation}
are the two distinguished  integrable systems of hydrodynamical
type (see e.g. \cite{dub,ben}). 1-layer Benney system describes
long waves in shallow water with free surface in gravitational
field. It is the dispersionless limit of the nonlinear
Schr\"{o}dinger equation \cite{zak}. Recently, the 1-layer Benney
($B(1)$) system became a crucial ingredient in the analysis of the
universality of critical behavior for nonlinear equations
\cite{dubgr}. The dToda equation is the 1+1-dimensional version of
the Boyer-Finley equation from the general relativity
\cite{boyer}. It shows up in various problems of fluid mechanics (
see e.g. \cite{mineev}). It is  known also (see e.g.\cite{dsl})
that the hodograph equations of the dToda hierarchy determine the
large $N$-limit of the Hermitian model in random matrix  theory.
In general, these two systems  are an excellent laboratory for
studying properties of integrable hydrodynamical type systems.

 In the present work  we analyze the structures  of the set of  hodograph equations of the  $B(1)$
 hierarchy and dToda hierarchy in terms of its Riemann invariants. These hodograph solutions describe the critical
points

\begin{equation}\label{crit}
\dfrac{\partial
W}{\partial\beta_i}=0,\quad i=1,2,
\end{equation}
of a function $W=W(\bt,\beta_1,\beta_2)$ which depend linearly on
the coordinates $\bt$, where $\bt$ denotes the flow parameters of
the Benney hierarchy and dToda hierarchy respectively and obey an
Euler-Poisson-Darboux equations $E(\varepsilon,\varepsilon)$
\cite{dar}
\begin{equation}\label{edpa}
(\beta_1\,-\,\beta_2)\, \dfrac{\partial^2
W}{\partial\beta_1\,\partial\beta_2}\,=\varepsilon\Big(\,\dfrac{\partial
W}{\partial\beta_1}\,-\,\dfrac{\partial W}{\partial\beta_2}\Big).
\end{equation}
where for the Benney system one has $\varepsilon=1/2$ and for the
dToda system $\varepsilon=-1/2$. The  equation \eqref{edpa} and
its multidimensional version are well known for a long time in
classical geometry \cite{dar}.
 Its relevance to the theory of Whitham equations has been demonstrated recently in the papers \cite{kud}-\cite{tian2}.
Here we will use classical notation $E(\varepsilon,\mu)$ for the
Euler-Poisson-Darboux equation proposed in \cite{dar} where such
equations with different $\varepsilon$ and $\mu$ have been studied
too.

If we denote by $\mathcal{M}$ the set of solutions $(\bt,\bb)\, (\beta_1\neq \beta_2)$ of the hodograph equations  \eqref{crit}, we may distinguish a regular and a singular sector in $\mathcal{M}$
\[
\mathcal{M}={\mathcal{M}}^{\mbox{reg}}\cup{\mathcal{M}}^{\mbox{sing}},
\]
such that given $(\bt,\bb)\in \mathcal{M}$
\[
\mbox{$(\bt,\bb)\in\mathcal{M}^{\mbox{reg}}$ if $\det \Big(\dfrac{\partial^2\,W(\bt,\bb)}{\partial\,\beta_i\,\partial\,\beta_j}\Big)\neq 0$},\quad
\mbox{$(\bt,\bb)\in\mathcal{M}^{\mbox{sing}}$ if $\det \Big(\dfrac{\partial^2\,W(\bt,\bb)}{\partial\,\beta_i\,\partial\,\beta_j}\Big)=0$}.
\]
 The elements of $\mathcal{M}^{\mbox{reg}}$, correspond to the case when the system \eqref{crit} is uniquely solvable and hence, it defines a unique solution $\bb(\bt)$.
The singular class  $\mathcal{M}^{\mbox{sing}}$  represents degenerate critical points of the function $W$ and are the points on  which the implicit solutions $\bb(\bt)$ of the  hodograph equations exhibit ``gradient catastrophe''
behaviour. As we will see in this paper, the Euler-Poisson-Darboux equation is of great help to analyze  the structure of  $\mathcal{M}^{\mbox{sing}}$.
As the
illustration of the general result a complete list of solutions of the 1-layer Benney hierarchy
from $\mathcal{M}^{\mbox{sing}}$ depending on two parameters is
presented.
 We also discuss the relation between Euler-Poisson-Darboux equations with opposite $a$ and Euler-Poisson-Darboux
equations for symmetries and densities of integrals of motion for integrable hydrodynamical type systems.

\section{1-layer Benney hierarchy and its singular sector}

The $B(1)$ system \eqref{ben} is a member of a dispersionless integrable hierarchy of deformations of the curve
(see e.g. \cite{kod,km}).
\begin{equation}\label{curve}
p^2=(\lambda-\beta_1)\,(\lambda-\beta_2).
\end{equation}
where $u=-(\beta _{1}+\beta _{2}),v=\frac{1}{4}(\beta _{1}-\beta
_{2})^{2}$. The  flows $\bb(\bt)$ are characterized by the
following condition: There exists a family of functions
$S(\lambda,\bt,\bb)$ satisfying
\begin{equation}\label{kdV}
\partial_{t_n}\, S(\lambda,\bt,\bb(\bt))=\Omega_n(\lambda,\bb(\bt)),
\quad n\geq 1.
\end{equation}
where
\begin{equation}
\Omega_n(\lambda,\bb)=
\Big(\dfrac{\lambda^n}{\sqrt{(\lambda-\beta_1)\,(\lambda-\beta_2)}}\Big)_{\oplus}\,\sqrt{(\lambda-\beta_1)\,(\lambda-\beta_2)}.
\end{equation}
where $\oplus$ denotes the standard projection on the positive
powers of  $\lambda$.Functions $S$ which satisfy \eqref{kdV} are
referred to as \emph{action functions}  in the theory of
dispersionless integrable systems (see e.g. \cite{kri}). Notice
that for $n=1$ \eqref{kdV} reads
\[
p=\dfrac{\partial S}{\partial x},\quad x:=t_1,
\]
so that the sytem \eqref{kdV} is equivalent to
\begin{equation}\label{kdvp}
\partial_{t_n} p=\,\partial_x\,\Omega_n,
\end{equation}
and, in terms of Riemann invariants $\bb$, it can be rewritten in
the hydrodynamical form
\begin{equation}\label{eqbeta}
\partial_{t_n}\,\beta_i=\Big(\dfrac{\lambda^n}{\sqrt{(\lambda-\beta_1)\,(\lambda-\beta_2)}}\Big)_{\oplus}\,\Big|_{\lambda=\beta_i}\,\partial_x\,\beta_i,\quad i=1,2.
\end{equation}
The $t_2$-flow of this hierarchy is the $B(1)$ system \eqref{ben}
($t:=t_2$)
\begin{equation}\label{benr}
\begin{cases}
\partial_{t}\,\beta_1=\dfrac{1}{2}\,(3\,\beta_1+\beta_2)\,\beta_{1\,x},\\ \\
\partial_{t}\,\beta_2=\dfrac{1}{2}\,(3\,\beta_2+\beta_1)\,\beta_{2\,x}.
\end{cases}
\end{equation}
For $v>0$ the $B(1)$ system is hyperbolic while for $v<0$ it is elliptic.

\vspace{0.3cm}
It was proved in \cite{kmm} that the system \eqref{crit} for  the critical points of the function
\begin{equation}\label{w1}
W(\bt,\bb)\,:=\,\oint_{\gamma}\dfrac{\d \lambda}{2\,i\,\pi}\,\dfrac{V(\lambda,\bt)}{\sqrt{(\lambda-\beta_1)\,(\lambda-\beta_2)}},
\end{equation}
where $V(\lambda,\bt)=\sum_{n\geq 1}t_n\,\lambda^n$, is a system of hodograph equations for the Benney hierarchy. Moreover, the action function for the corresponding solutions is given by
\begin{equation}\label{sol}
S(\lambda,\bt,\bb)=\sum_{n\geq 1}\, t_n\,\Omega_n(\lambda,\bb)=h(\lambda,\bt,\bb)\,\sqrt{(\lambda-\beta_1)(\lambda-\beta_2)}.
\end{equation}
where
\[
h(\lambda,\bt,\bb):=\Big(\dfrac{V(\lambda,\bt)}{\sqrt{(\lambda-\beta_1)(\lambda-\beta_2)}}\Big)_{\oplus}.
\]
Obviously, the function $W$ satisfies the Euler-Poisson-Darboux equation $E(1/2, 1/2)$.

 Written explicitly, $W$ represents itself the series
\begin{align}\label{Wexp}
 W&=\,\dfrac{x}{2}(\beta_1+\beta_2)+\dfrac{t_2}{8}(3\beta_1^2+2\beta_1\beta_2+3\beta_2^2)+\dfrac{t_3}{16}
 \left(5\beta_1^3+
  3\beta_1^2\beta_2+3\beta_1\beta_2^2+5\beta_2^3\right) \nonumber \\
  &+
  \dfrac{t_4}{128}(35\beta_1^4+20\beta_1^3\beta_2+18\beta_1^2\beta_2^2+20\beta_1\beta_2^3+35\beta_2^4)+\cdots.
  \end{align}
 The hodograph equations \eqref{crit} with $t_n\,=\,0$ for $n\,\geq\,5$ take the form
 \begin{equation}\label{h2}
\everymath{\displaystyle}
\begin{cases}
8x+4t_2(3\beta_1+\beta_2)+3t_3\left(5\beta_1^2+
  2\beta_1\beta_2+\beta_2^2\right)+
  \dfrac{t_4}{8}(140\beta_1^3+60\beta_1^2\beta_2+18\beta_1\beta_2^2+20\beta_2^3)=0,\\\\
8x+4t_2(\beta_1+3\beta_2)+3t_3\left(\beta_1^2+
  2\beta_1\beta_2+5\beta_2^2\right)
  +\dfrac{t_4}{8}(140\beta_2^3+60\beta_2^2\beta_1+18\beta_2\beta_1^2+20\beta_1^3)=0.
\end{cases}
\end{equation}
Detailed analysis of equations \eqref{h2} will be performed in section 3.
Here, we would like to make two observations. First, one is that the formulae \eqref{h2} point out on the possible
alternative interpretation of the times $t_2$, $t_3$, $t_4$,... of the $B(1)$ hierarchy. Namely, taking $t_2\,=\,0$ in the
formulae \eqref{h2}, we see that $t_3$ and $t_4$ are parameters appearing in the initial data $\beta_1(x,t_2=0)$ and
$\beta_2(x,t_2=0)$. Thus, one can view hodograph equations \eqref{crit} (in particular, equations \eqref{h2}) as
equations describing the time evolution of the family of initial data for the $B(1)$ system , parametrized by the
variables $t_3$, $t_4$, $t_5$,...

Second observation concerns with the elliptic version of the $B(1)$ system. In this case
$\beta_2\,=\,\overline{\beta}_1$ and the system \eqref{benr} reduces to the single equation
\begin{equation}\label{redben}
\partial_{t}\,\beta\,=\,\frac{1}{2}\,(3\,\beta\,+\,\overline{\beta})\,\beta_x,\quad t:=t_2,\quad \beta:=\beta_1.
\end{equation}
This equation is equivalent to the nonlinear Beltrami equation
\begin{equation}\label{bel}
\beta_{\bar{z}}\,=\,\frac{2\,i\,-\,3\,\beta\,-\,\overline{\beta}}{2\,i\,+\,3\,\beta\,+\,\overline{\beta}}\,\beta_z,
\end{equation}
where $z\,=\,x\,+\,i\,t$. This fact indicates that the theory of quasi-conformal mappings (see e.g. \cite{alp})
can be relevant for the analysis of properties of the elliptic $B(1)$ system \eqref{ben} $(v\,<\,0)$. Hence,
since the elliptic $B(1)$ system is the quasiclassical limit  \cite{zak} of the focusing nonlinear Schr\"{o}dinger (NLS) equation
\[
i\,\epsilon\,\psi_t\,+\,\dfrac{\epsilon^2}{2}\,\psi_{xx}\,+\,|\psi|^2
\,\psi\,=\,0,
\]
with $\psi\,=\,A\,\exp{\Big(\dfrac{i}{\epsilon}\,S\Big)}$,
$u\,=\,\dfrac{\epsilon}{2\,i}\Big(\dfrac{\psi_x}{\psi}-\dfrac{\overline{\psi}_x}{\overline{\psi}}\Big)$, $v\,=\,-\,|\psi|^2$,
$\epsilon\,\rightarrow\,0$, the quasiconformal mapping can be useful also in the study of the small dispersion limit of the focusing
NLS equation (compare with \cite{dubgr}).

In order to analyse singular sector of the 1-layer Benney
hierarchy we first observe that due to the Euler-Poisson-Darboux
equation for given $(\bt,\bb)\in \mathcal{M}$,  as a consequence
of \eqref{edpa} one has
 \begin{equation}\label{nondiag}
\frac{\partial^2\,W}{\partial\,\beta_1\,\partial\,\beta_2}\,=\,0.
\end{equation}
Consequently
\begin{equation}\label{det}
\det\Big(\frac{\partial^2\,W}{\partial\,\beta_i\,\partial\,\beta_j}\Big)=
\frac{\partial^2\,W}{\partial\,\beta_1^2}\cdot\frac{\partial^2\,W}{\partial\,\beta_2^2}.
\end{equation}
Thus, we have
\begin{pro}
Given $(\bt,\bb)\in \mathcal{M}$ then
\begin{enumerate}
\item $(\bt,\bb)\,\in\,\mathcal{M}^{\mbox{reg}}$ if and only if
$\everymath{\displaystyle}
\frac{\partial^2\,W}{\partial\,\beta_1^2}\,\neq\,0\, \mbox{and} \,
\frac{\partial^2\,W}{\partial\,\beta_2^2}\,\neq\,0.
$
\item $(\bt,\bb)\,\in\,\mathcal{M}^{\mbox{sing}}$ if and only at least one of the
derivatives
$\everymath{\displaystyle}
\frac{\partial^2\,W}{\partial\,\beta_1^2},\, \frac{\partial^2\,W}{\partial\,\beta_2^2},
$
vanishes.
\end{enumerate}
\end{pro}
Furthermore, using \eqref{edpa} it follows easily that   at any point $(\bt,\bb)\in {\mathcal{M}}$ all mixed derivatives $\partial_{\beta_1}^i\partial_{\beta_2}^j W$ can be expressed in terms of linear combination of derivatives  $\partial_{\beta_1}^nW$ and  $\partial_{\beta_2}^m W$. Hence  if we define ${\mathcal{M}}^{\mbox{sing}}_{n_1,n_2}$ as the set of points $(\bt,\bb)\in {\mathcal{M}}$ such that
\begin{equation}\label{m12}
\dfrac{\partial^{n_i+2}W}{\partial\beta_i^{n_i+2}}\neq 0, \quad
\dfrac{\partial^k W}{\partial \beta_i^k}=0,\quad  \forall 1\leq k\leq n_i+1,\quad (i=1,2),
\end{equation}
it follows that
$$
{\mathcal{M}}^{\mbox{sing}}=\bigcup _{n_1+n_2 \geq 1}{\mathcal{M}}^{\mbox{sing}}_{n_1,n_2},
$$
where
$$
{\mathcal{M}}^{\mbox{sing}}_{n_1,n_2}\bigcap{\mathcal{M}}^{\mbox{sing}}_{n'_1,n'_2}=\emptyset,\;
\mbox{ for $(n_1,n_2)\neq (n'_1,n'_2)$}
$$

We may  characterize  the  classes $\mathcal{M}^{\mbox{sing}}_{n_1,n_2}$ of the singular sector in terms of the behaviour of $S(\lambda)$
at  $\lambda=\beta_i\, (i=1,2)$ . Indeed the derivative $\partial_{\beta_1}^{k+1}W$ with $k\geq 1$ is proportional to the integral
\begin{equation*}\everymath{\displaystyle}
\oint_{\gamma}\dfrac{\d \lambda}{2\,i\,\pi}
\,\dfrac{V(\lambda,\bt)}{
(\lambda-\beta_1)^{k+1}\,\sqrt{(\lambda-\beta_1)\,(\lambda-\beta_2)}}=\oint_{\gamma}\dfrac{\d \lambda}{2\,i\,\pi}\
\dfrac{h(\lambda)}{
(\lambda-\beta_1)^{k+1}}
=\dfrac{1}{k!}\,\Big(\partial_{\lambda}^{k} \,h(\lambda)\Big)\Big|_{\lambda=\beta_1},
\end{equation*}
and a similar result follows for the derivatives  $\partial_{\beta_2}^{k+1}W$ with $k\geq 2$.  As a consequence we have
\begin{pro}
A point  $(\bt,\bb)\in\mathcal{M}$ belongs to the singularity class $\mathcal{M}^{\mbox{sing}}_{n_1,n_2}$ if and only if
\begin{equation}
\everymath{\displaystyle}
S(\lambda,\bt,\bb)\sim (\lambda-\beta_i)^{\frac{2n_i+3}{2}}\quad \mbox{as $\lambda\rightarrow \beta_i,\quad (i=1,2)$}
\end{equation}
\end{pro}
\noindent

\section{Explicit determination of singular sectors}

It is easy  to see that the singular classes  ${\mathcal{M}}^{\mbox{sing}}_{n_1,n_2}$ can be determined by means of a system of $n_1+n_2$ constraints for the coordinates $\bt$. Indeed, the points $(\bt,\bb)$ of  $ {\mathcal{M}}^{\mbox{sing}}_{n_1,n_2}$ are characterized by the equations
\begin{equation}\label{m12b}
\dfrac{\partial^k W}{\partial \beta_i^k}=0,\quad  \forall 1\leq k\leq n_i+1,\quad i=1,2,
\end{equation}
and
\begin{equation}\label{m12c}
\dfrac{\partial^{n_i+2}W}{\partial\beta_i^{n_i+2}}\neq 0, \quad i=1,2.
\end{equation}
Now the observation is that the jacobian matrix of the the system of two equations
\begin{equation}\label{m12d}
\dfrac{\partial^{n_i+1}W}{\partial\beta_i^{n_i+1}}=0, \quad i=1,2
\end{equation}
is not singular as
\begin{equation}\label{Delta23}\everymath{\displaystyle}
\Delta\,:=\,\left|\begin{array}{cc}
\frac{\partial^{n_1+2}\,W}{\partial\,\beta_1^{n_1+2}} & \frac{\partial^{n_2+2}\,W}{\partial\, \beta_1\partial\,\beta_2^{n_2+1}}
 \\  \\
 \frac{\partial^{n_1+2}\,W}{\partial\,\beta_1^{n_1+1}\,\partial\,\beta_2} & \frac{\partial^{n_2+2}\,W}{\partial\,\beta_2^{n_2+2}}
\end{array}\right|\,\neq\,0.
\end{equation}
Indeed, we notice that as a consequence of \eqref{edpa} the derivatives outside the diagonal of $\Delta$ are  linear combinations of  the derivatives  $\{\partial_{\beta_i}^k\,W, \; 1\leq k\leq n_i+1,\; i=1,2\}$, so that from \eqref{m12b}-\eqref{m12c} we have
\[
\Delta=\frac{\partial^{n_1+2}\,W}{\partial\,\beta_1^{n_1+2}}\cdot \frac{\partial^{n_2+2}\,W}{\partial\,\beta_2^{n_2+2}}\neq 0.
\]
Therefore, one can solve \eqref{m12d} and get a solution $\bb(\bt)$. Substituting this solution in the remaining equations
\eqref{m12b} gives $n_1+n_2$ constraints of the form
\[
f_k(\bt)=0,\quad k=1,\ldots,n_1+n_2.
\]
It is not difficult to determine the solutions  of \eqref{m12b}-\eqref{m12c} in two simple cases: with one parameter $t_3$ ($t_4=t_5=\cdots=0$),
and with two parameters $t_3$, $t_4$ ($t_5=t_6=\cdots=0$). We have that in this case
\[
{\mathcal{M}}^{\mbox{sing}}\,=\,{\mathcal{M}}^{\mbox{sing}}_{10}\,\cup\,{\mathcal{M}}^{\mbox{sing}}_{01}
\]
with ${\mathcal{M}}^{\mbox{sing}}_{10}$ defined by
$$\everymath{\displaystyle}\begin{array}{ll}
\textbf{1.} &
x\,=\,\frac{-45t_4 t_3^3+180 t_2 t_4^2
   t_3+\sqrt{15}(8t_2t_4-3t_3^2)\sqrt{t_4^2 \left(3 t_3^2-8
   t_2 t_4\right)}}{360 t_4^3},\\  \\
   &\beta_1\,=\,-\frac{5 t_3 t_4+\sqrt{15} \sqrt{t_4^2 \left(3
   t_3^2-8 t_2 t_4\right)}}{20t_4^2},\qquad \beta_2\,=\,\frac{-3 t_3 t_4+\sqrt{15} \sqrt{t_3^2 \left(3 t_3^2-8 t_2
   t_4\right)}}{12 t_4^2},\\  \\  \\
\textbf{2.} & x\,=\,\frac{-45t_4 t_3^3+180 t_2 t_3^2
   t_3-\sqrt{15}(8t_2t_4-3t_3^2)\sqrt{t_4^2 \left(3 t_3^2-8
   t_2 t_4\right)}}{360 t_4^3},\\  \\
   &\beta_1\,=\,\frac{-5 t_3 t_4+\sqrt{15} \sqrt{t_4^2 \left(3
   t_3^2-8 t_2 t_4\right)}}{20t_4^2},\qquad \beta_2\,=\,-\frac{3 t_3 t_4+\sqrt{15} \sqrt{t_4^2 \left(3 t_3^2-8 t_2
   t_4\right)}}{12 t_4^2},
\end{array}$$
and ${\mathcal{M}}^{\mbox{sing}}_{01}$ by
$$\everymath{\displaystyle}\begin{array}{ll}
\textbf{3.} & x\,=\,\frac{-45t_4 t_3^3+180 t_2 t_4^2
   t_3-\sqrt{15}(8t_2t_4-3t_3^2)\sqrt{t_4^2 \left(3 t_3^2-8
   t_2t_4\right)}}{360 t_4^3},\\  \\
   &\beta_1\,=\,-\frac{3 t_3 t_4+\sqrt{15} \sqrt{t_4^2 \left(3 t_3^2-8 t_2
   t_4\right)}}{12 t_4^2},\qquad \beta_2\,=\,\frac{-5 t_3 t_4+\sqrt{15} \sqrt{t_4^2 \left(3
   t_3^2-8 t_2 t_4\right)}}{20t_4^2},\\  \\  \\
\textbf{4.} & x\,=\,\frac{-45t_4 t_3^3+180 t_2 t_4^2
   t_3+\sqrt{15}(8t_2t_4-3t_3^2)\sqrt{t_4^2 \left(3 t_3^2-8
   t_2 t_4\right)}}{360 t_4^3},\\  \\
   &\beta_1\,=\,\frac{-3 t_3 t_4+\sqrt{15} \sqrt{t_4^2 \left(3 t_3^2-8 t_2
   t_4\right)}}{12 t_4^2},\qquad \beta_2\,=\,-\frac{5 t_3 t_4+\sqrt{15} \sqrt{t_4^2 \left(3
   t_3^2-8 t_2 t_4\right)}}{20t_4^2}.
   \end{array}$$

\section{Singular sector of the elliptic $B(1)$ system}

Now, we will consider the elliptic $B(1)$ system \eqref{ben}. Singular sector ${\mathcal{M}}^{\mbox{sing}}$ has
in this case a structure which is quite different from that of the hyperbolic system. Indeed, since $\beta_1\,=\,\overline{\beta}_2$,
the function $W$ for real $x,\,t_2,\,t_3,\,\dots$ is the real valued function
\[
\overline{W(\bt,\beta,\overline{\beta})}\,=\,W(\bt,\beta,\overline{\beta}),
\]
and the hodograph equation \eqref{crit} has the form of the Cauchy-Riemann condition $(\beta\,=\,\beta_1)$
\begin{equation}\label{CR}
\frac{\partial\,W}{\partial\,\overline{\beta}}\,=\,0.
\end{equation}

Regular and singular sectors $\mathcal{M}^{\mbox{reg}}$ and $\mathcal{M}^{\mbox{sing}}$ are defined as the sets
$(\bt,\beta,\overline{\beta})$ of solutions of equation \eqref{CR} such that the hermitian form
$$\d^2\,W\,=\,\frac{\partial^2\,W}{\partial\,{\beta}^2}\,\d\beta^2\,+\,
2\,\frac{\partial^2\,W}{\partial\,\beta\,\partial\,\overline{\beta}}\,\d\beta\,\d\,\overline{\beta}\,+\,
\frac{\partial^2\,W}{\partial\,\overline{\beta}^2}\,\d\,\overline{\beta}^2,$$
is nondegenerate or degenerate, respectively. For unreduced solutions $(\beta\,\neq\,\overline{\beta})$, the corresponding
Euler-Poisson-Darboux equation implies that
$$\frac{\partial^2\,W}{\partial\,\beta\,\partial\,\overline{\beta}}\,=\,0,$$
and, hence
\begin{equation}\label{det}
\everymath{\displaystyle}\left|
\begin{array}{cc}
\frac{\partial^2\,W}{\partial\,{\beta}^2} &
\frac{\partial^2\,W}{\partial\,\beta\,\partial\,\overline{\beta}}\\  \\
\frac{\partial^2\,W}{\partial\,\beta\,\partial\,\overline{\beta}} &
\frac{\partial^2\,W}{\partial\,\overline{\beta}^2}
\end{array}\right|\,=\,
\left|\frac{\partial^2\,W}{\partial\,\overline{\beta}^2}\right|^2.
\end{equation}
Thus, one has

\begin{pro}
For unreduced solutions of the elliptic $B(1)$ system, the regular sector $\mathcal{M}^{\mbox{reg}}$ is defined by the condition
\begin{equation}\label{regconel}
\frac{\partial\,W}{\partial\,\overline{\beta}}\,=\,0,\quad \frac{\partial^2\,W}{\partial\,\overline{\beta}^2}\,\neq\,0.
\end{equation}
\end{pro}

A similar analysis to that of the hyperbolic case readily leads to

\begin{pro}
Singular sector $\mathcal{M}^{\mbox{sing}}$ of the elliptic $B(1)$ system \eqref{ben} is the union of the subspaces
$\mathcal{M}_n^{\mbox{sing}}$, $(n\,=\,1,2,3,\dots)$ defined as
\begin{equation}\label{singellip}
\mathcal{M}_n^{\mbox{sing}}\,=\,\left\{(\bt,\beta,\overline{\beta})\,\in\,\mathcal{M}^{\mbox{sing}}:\;
\frac{\partial^k\,W}{\partial\,\overline{\beta}^k}\,=\,0,\;k\,=\,1,\dots,n+1;\;
\frac{\partial^{(n+2)}\,W}{\partial\,\overline{\beta}^{(n+2)}}\,\neq\,0\right\}
\end{equation}
Solutions belonging to $\mathcal{M}_n^{\mbox{sing}}$ are defined on a subspace of codimension $2\,n$ in the space of parameters
$x$, $t_2$, $t_3$, ... .
\end{pro}

So, in the elliptic case, gradient catastrophe happens in the point $(x,t)$ at fixed parameters $t_2$, $t_3$,...
Similar to the hyperbolic case, the subspace $\mathcal{M}_n^{\mbox{sing}}$ is not empty if at least $n$ parameters
$t_2$, $t_3$,...,$t_{n+1}$ are different from zero in the formula \eqref{Wexp}.

It is instructive to rewrite the formula \eqref{Wexp} for the function $W$ in terms of the real and imaginary part of $\beta_1$. i.e.
$\beta_1\,=\,U\,+\,i\,V$:
\begin{equation}\label{Wexpri}\everymath{\displaystyle}\begin{array}{lll}
W&=&x\,U\,+\,t_2\,(U^2\,-\,\frac{1}{2}V^2)\,+\,t_3\,(U^3\,-\,\frac{3}{2}\,U\,V^2)\,+\,
t_4\,(U^4\,-\,3\,U^2\,V^2\,+\,\frac{3}{8}\,V^4)\\  \\
 & &\,+\,t_5\,(U^5\,-\,5\,U^3\,V^2\,+\,\frac{15}{8}\,U\,V^4)\,+\,\cdots.
\end{array}\end{equation}
This formula explicitly shows the character of elliptic singularities exhibited for the function $W$ for various values of
parameters $t_2$, $t_3$, ....

Basic equations \eqref{CR}, \eqref{Wexp} and also conditions \eqref{singellip} defining subspaces $\mathcal{M}_n^{\mbox{sing}}$
can be easily rewritten in terms of the original variables $u$ and $v$. Since
$$\frac{\partial\,W}{\partial\,\beta}\,=\,-\,\frac{\partial\,W}{\partial\,u}\,+\,i\,\sqrt{-\,v}\,\frac{\partial\,W}{\partial\,v},$$
the hodograph equation \eqref{CR} becomes (for $v\,\neq\,0$)
\begin{equation}\label{conuv}
\frac{\partial\,W}{\partial\,u}\,=\,0,\quad \frac{\partial\,W}{\partial\,v}\,=\,0,
\end{equation}
while the Euler-Poisson-Darboux equation and equation \eqref{conuv} take the form
\begin{equation}\label{epdellip}
\frac{\partial^2\,W}{\partial\,u^2}\,-\,v\,\frac{\partial^2\,W}{\partial\,v^2}\,=\,0.
\end{equation}
For the subspace $\mathcal{M}_1^{\mbox{sing}}$ conditions \eqref{singellip} are
\begin{equation}\label{s1e}
\frac{\partial\,W}{\partial\,\beta}\,=\,0,\quad \frac{\partial^2\,W}{\partial\,\beta^2}\,=\,0,\quad
\frac{\partial^3\,W}{\partial\,\beta^3}\,\neq\,0.
\end{equation}
Since
$$\frac{\partial^2\,W}{\partial\,\beta^2}\,=\,\frac{\partial^2\,W}{\partial\,u^2}\,-\,2\,i\,\sqrt{-\,v}\,
\frac{\partial^2\,W}{\partial\,u\,\partial\,v}\,+\,v\,\frac{\partial^2\,W}{\partial\,v^2}\,+\,
\frac{1}{2}\,\frac{\partial\,W}{\partial\,v},$$
one concludes taking into account equation \eqref{conuv} and \eqref{epdellip} that the second condition \eqref{s1e}
is satisfied if and only if
\begin{equation}\label{ss}
\frac{\partial^2\,W}{\partial\,u^2}\,=\,0,\quad \frac{\partial^2\,W}{\partial\,v^2}\,=\,0,\quad
\frac{\partial^2\,W}{\partial\,u\,\partial\,v}\,=\,0.
\end{equation}
Thus, the subspace $\mathcal{M}_1^{\mbox{sing}}$ is characterized by the conditions \eqref{conuv}, \eqref{ss} and by requirement of
nonvanishing third order derivatives of $W$.

In order to compare these conditions with those of paper \cite{dubgr}, we first observe that the $B(1)$ system \eqref{ben}
is converted into the system (1.8) by the substitution $u\,\rightarrow\,v$, $v\,\rightarrow\,-\,u$. Then, with the choice
$$W\,=\,f(u,v)\,+\,x\,v\,-\,u\,v\,t,$$
the hodograph equations \eqref{conuv} become equations (2.4) of \cite{dubgr} and equation \eqref{epdellip} is reduced to
their equation (2.5). Finally, with such a choice, the conditions \eqref{ss} are converted to the condition (2.12) from
the paper \cite{dubgr}.

Finally, we note that according to the proposition 4 for the subspace $\mathcal{M}_1^{\mbox{sing}}$, the codimension of the
corresponding subspace of $(x,t_2,t_3,\dots)$ is equal to two and the function $W$ with $t_n\,=\,0$, $n\,\geq\,4$ i.e.
$$W\,=\,x\,U\,+\,t_2\,(U^2\,-\,\frac{1}{8}V^2)\,+\,t_3\,(U^3\,-\,\frac{3}{2}\,U\,V^2),$$
exhibits the elliptic umbilic singularity according to Thom's classification \cite{thom} (see also \cite{alp}-\cite{arn2}).
These results reproduce those originally obtained in the paper \cite{dubgr} (formula (4.2))

\section{5. dToda hierarchy.}

 Now let us consider the function
\begin{equation}
W_{T}(x,\beta _{1,}\beta _{2})=\int \frac{d\lambda }{2\pi i}\,V_{T}(x,\lambda )%
\sqrt{(1-\frac{\beta _{1}}{\lambda })\,(1-\frac{\beta _{2}}{\lambda })}
\end{equation}
where $V_{T}(x,\lambda )=\sum_{n\geq 0}\lambda ^{n}x_{n}.$ Critical points for this
function are defined by the equations
\begin{equation}
\frac{\partial W_{T}}{\partial \beta _{1}}=0,\quad \frac{\partial
W_{T}}{\partial \beta _{2}}=0.
\end{equation}
It is a simple check to see that $W_{T}$ obeys the Euler-Poisson-Darboux equation
of the type $E(-1/2,-1/2)$
\begin{equation}
2(\beta _{1}-\beta _{2})\frac{\partial ^{2}W_{T}}{\partial \beta
_{1}\partial \beta _{2}}=-(\frac{\partial W_{T}}{\partial \beta _{1}}-\frac{%
\partial W_{T}}{\partial \beta _{2}}).
\end{equation}
Written explicitly the function $W_{T}$ is the series
\begin{equation}
W_{T}=-\frac{1}{2}x_{0}(\beta _{1}+\beta _{2})-\frac{1}{8}x_{1}(\beta
_{1}-\beta _{2})^{2}-\frac{1}{16}x_{2}(\beta _{1}+\beta _{2})(\beta
_{1}-\beta _{2})^{2}-\frac{1}{128}x_{3}(5\beta _{1}^{2}+6\beta _{1}\beta
_{2}+5\beta _{1}^{2})(\beta _{1}-\beta _{2})^{2}+...
\end{equation}
while the hodograph equations take the form
\begin{align*}
& x_{0}+\frac{1}{2}x_{1}(\beta _{1}-\beta _{2})+\frac{1}{8}x_{2}(3\beta
_{1}^{2}-2\beta _{1}\beta _{2}-\beta _{2}^{2})+\ldots=0,  \\
&x_{0}-\frac{1}{2}x_{1}(\beta _{1}-\beta
_{2})+\frac{1}{8}x_{2}(3\beta _{2}^{2}-2\beta _{1}\beta _{2}-\beta
_{1}^{2})+\ldots=0.
\end{align*}
These hodograph equations provide us with the solutions of the system
\begin{equation}
\frac{\partial \beta _{1}}{\partial x_{1}}=\frac{1}{2}(\beta _{1}-\beta _{2})%
\frac{\partial \beta _{1}}{\partial x_{0}},\quad \frac{\partial \beta _{2}}{%
\partial x_{1}}=-\frac{1}{2}(\beta _{1}-\beta _{2})\frac{\partial \beta _{2}%
}{\partial x_{0}}.
\end{equation}
In terms of the variables $u=-(\beta _{1}+\beta _{2}),v=\frac{1}{4}(\beta
_{1}-\beta _{2})^{2}$ one has the dToda system (2). \ Considering the higher
times $x_{2},x_{3},...$ .one gets the whole dToda hierarchy.

Similar to the Benney case \ the function $W_{T}$ is the generating function
for classical singularities for functions of two variables. Indeed,
in the variables $X=\frac{1}{2}(\beta _{1}+\beta _{2}),Y=\frac{1}{2}(\beta
_{1}-\beta _{2})$ it is of the form
\begin{equation}
W_{T}=-x_{0}X-\frac{1}{2}x_{1}Y^{2}-\frac{1}{2}x_{2}XY^{2}-\frac{1}{8}x_3
(4X^{2}+Y^{2})Y^{2}+...
\end{equation}
The third term here represents the parabolic umbilic singularity both for
hyperbolic and elliptic cases.

The formulas for the dToda hierarchy presented here coincide with
those given in the paper \cite{konopel} after the identification
\begin{equation}
V_{T}(x,\lambda )=-2T\lambda +\lambda V_{H}^{\prime }(t,\lambda ).
\end{equation}
i.e. $x_{0}=-2T,x_{n}=nt_{n},n=1,2,3,...$.

It is obvious that the descriptions of the regular and singular sectors of
the dToda hierarchy completely coincide with those of 1-layer Benney
hierarchy.

\bigskip

\section{6. Interrelations between the Euler-Poisson-Darboux equations with different indices
 and those for function W and densities of integrals of motion.}

\bigskip

1-layer Benney hierarchy and dToda hierarchy are two examples of
hydrodynamical type systems for which functions $W$ obey the
Euler-Poisson-Darboux equations
\begin{equation}
L_{\varepsilon }W_{\varepsilon }:= \Big[\frac{\partial ^{2}}{\partial
x\partial y}-\dfrac{\varepsilon}{x-y} \Big(\frac{\partial }{\partial x}-\frac{\partial }{%
\partial y}\Big)\Big]W_{\varepsilon }=0,
\end{equation}
with different indexes $\varepsilon $. \ Such linear equations \ \ are well
studied ( see e.g. \cite{dar}). The operators $L_{\varepsilon }$ have a number
of remarkable properties. One of them ( probably missed before) is given by the identity
\begin{equation}
L_{\varepsilon +1}L_{\mu }=L_{\mu +1}L_{\varepsilon }
\end{equation}
for arbitrary indices $\varepsilon$ and $\mu$ . This identity
implies, for instance, that for any solution $W_{\varepsilon}$  the function $L_{\mu}W_{\varepsilon}$ with arbitrary
$\mu$ obeys the
Euler-Poisson-Darboux equation with index $\varepsilon +1$, more precisely $%
L_{\mu }W_{\varepsilon }=\varepsilon (\varepsilon -\mu )W_{\varepsilon +1}$
. In particular, at $\varepsilon =-\frac{1}{2}$ and $\mu =0$ one has $L_{%
\frac{1}{2}}L_{0}=L_{1}L_{-\frac{1}{2}}.$ In terms of the operators $%
\widetilde{L}_{\varepsilon }$ defined as $\widetilde{L}_{\varepsilon
}=(x-y)L_{\varepsilon}$ the last relation takes the form
\begin{equation}
\partial _{x}\partial _{y}\widetilde{L}_{-\frac{1}{2}}=\widetilde{L}_{\frac{1%
}{2}}\partial _{x}\partial _{y}.
\end{equation}
This identity clearly demonstrates the duality between the
Euler-Poisson-Darboux equations with indices $\frac{1}{2}$ and -$\frac{1}{2}$
and consequently between 1-layer Benney \ and dToda hierarchies.

Duality between the functions W and densities of integrals of motions is the
another type of duality typical for the so-called $\varepsilon$ integrable
hydrodynamical type systems. Indeed, due to the Tsarev's result \cite{tsa} , a
symmetry $w_{i}$ of a semi-Hamiltonian hydrodynamical system
\begin{equation}\label{44}
\frac{\partial \beta _{i}}{\partial t}=\lambda _{i}(\beta )\frac{\partial
\beta _{i}}{\partial x},\quad i=1,...,n,
\end{equation}
i.e. a solution of the system
\begin{equation}
\frac{\partial \beta _{i}}{\partial \tau }=w_{i}(\beta )\frac{\partial \beta
_{i}}{\partial x},\quad i=1,...,n
\end{equation}
which commutes with the system \eqref{44}, are defined by the system
\begin{equation}\everymath{\displaystyle}
\frac{\frac{\partial w_{k}}{\partial \beta _{i}}}{w_{i}-w_{k}}=\frac{\frac{
\partial \lambda _{k}}{\partial \beta _{i}}}{\lambda _{i}-\lambda _{k}},\quad i\neq k.
\end{equation}
Such $w_{i}$ provide us with the solutions of the systems \eqref{44} via
the hodograph equations
\begin{equation}
\Omega _{i}:= -x+\lambda _{i}(\beta )t+w_{i}=0,i=1,...,n.
\end{equation}
For such system \eqref{44} densities P of integrals of motion obey  the
equations \cite{tsa}

\begin{equation}
\frac{\partial ^{2}P}{\partial \beta _{i}\partial \beta _{k}}=\frac{\frac{%
\partial \lambda _{i}}{\partial \beta _{k}}}{\lambda _{i}-\lambda _{k}}\frac{%
\partial P}{\partial \beta _{i}}-\frac{\frac{\partial \lambda _{k}}{\partial
\beta _{i}}}{\lambda _{i}-\lambda _{k}}\frac{\partial P}{\partial \beta _{k}}
,\quad i\neq k.
\end{equation}%
Let us define $\varepsilon $-systems as those ( for particular class of such
systems see e.g. \cite{pav2003}) for which
\begin{equation}
\frac{\frac{\partial \lambda _{i}}{\partial \beta _{k}}}{\lambda
_{i}-\lambda _{k}}=\frac{\frac{\partial \lambda _{k}}{\partial \beta _{i}}}{%
\lambda _{i}-\lambda _{k}}=\frac{\varepsilon }{\beta _{i}-\beta _{k}}
\end{equation}%
For such systems densities of integrals obey
Euler-Poisson-Darboux equations
\begin{equation}
\frac{\partial ^{2}P}{\partial \beta _{i}\partial \beta _{k}}=\frac{%
\varepsilon }{\beta _{i}-\beta _{k}}\frac{\partial P}{\partial \beta _{i}}-%
\frac{\varepsilon }{\beta _{i}-\beta _{k}}\frac{\partial P}{\partial \beta
_{k}},\quad i\neq k.
\end{equation}%
At the same time the equations for $w_{i}$ become
\begin{equation}
\frac{\partial w_{k}}{\partial \beta _{i}}=-\varepsilon \frac{w_{i}-w_{k}}{%
\beta _{i}-\beta _{k}},\quad i\neq k
\end{equation}%
Symmetry of these equations with respect to the transposition of indices $i$ and
$k$ implies that
$\frac{\partial w_{k}}{\partial \beta _{i}}=\frac{\partial
w_{i}}{\partial \beta k}.$ Hence
$$
w_{i}=\frac{\partial \widetilde{W}}{%
\partial \beta _{i}},\quad i=1,...,n,
$$
for a certain function $\widetilde{W}$.
Thus, equations (51) are the Euler-Poisson-Darboux equations of the type $%
E(-\varepsilon ,-\varepsilon )$ for the function $\widetilde{W}$
\begin{equation}
\frac{\partial ^{2}\widetilde{W}}{\partial \beta _{i}\partial \beta _{k}}=-\Big(%
\frac{\varepsilon }{\beta _{i}-\beta _{k}}\frac{\partial \widetilde{W}}{%
\partial \beta _{i}}-\frac{\varepsilon }{\beta _{i}-\beta _{k}}\frac{%
\partial \widetilde{W}}{\partial \beta _{k}}\Big),\quad i\neq k.
\end{equation}%
The fact that the generating function for symmetries of the Whitham
equations and some other integrable hydrodynamical systems obey the
Euler-Poisson-Darboux equations has been observed earlier in the papers \cite{kud,pav,pav2003}.
Also the duality between the Euler-Poisson-Darboux equations for the densities of integrals
of motions and generating functions of symmetries has been noted before too. However the
demonstration presented above seems to be different from those discussed earlier.

In addition one can note that equations (49) imply that for $\varepsilon $%
-systems also $\lambda _{i}=\frac{\partial g}{\partial \beta _{i}}$ with
some function $g$. \ As the result the hodograph equations (47) for the $%
\varepsilon $-systems take the form
\begin{equation}
\Omega _{i}=-x+t\frac{\partial g}{\partial \beta _{i}}+\frac{\partial
\widetilde{W}}{\partial \beta _{i}}=\frac{\partial W}{\partial \beta _{i}}%
=0,i,...,n
\end{equation}
where $W=-x(\beta _{1}+\beta _{2})+gt+\widetilde{W.}$ Thus,
hodograph equations for the integrable hydrodynamical type systems
\ are nothing but the equations defining the critical points of
the function $W$. \ It seems \ that this fact has been missing in
the previous publications. Moreover, due to the equations (49) the
function $g$ also obeys the $E(-\varepsilon ,-\varepsilon )$
Euler-Poisson-Darboux equation and , hence, the function $W$ does
the same. Note that particular class of $\varepsilon $-systems for
which $\lambda _{i}$ are linear functions of $\beta_i$ has been
discussed in \cite{pav2003}.

  So, for integrable hydrodynamical systems of the
$\varepsilon $ type,  the densities of integrals and the functions
$W$ ( as well as the functions $\widetilde{W}$ generating
symmetries) play a dual role
obeying the Euler-Poisson-Darboux equations with opposite sign of the index $%
\varepsilon $ . This property resembles a lot the well-known duality between
the generating functions of integrals of motion and symmetries for the
dispersionful integrable equations.

\vspace{0.5cm}

\subsection* { Acknowledgements}

\vspace{0.3cm} The authors  wish to thank the  Spanish Ministerio de
Educaci\'on y Ciencia (research project FIS2008-00200/FIS).

\end{document}